\documentclass[slac_one]{revtex4}
\usepackage{graphicx}
\usepackage{fancyhdr}
\usepackage{amsfonts}
\begin{document}

\title{Improving the precision of $\gamma/\phi_3$ via CLEO-c Measurements} 

%

\author{
P. Naik (for the CLEO collaboration)
}
\affiliation{University of Bristol, Bristol BS8 1TL, UK}
%

\begin{abstract}
Quantum correlations in $\psi(3770) \to D^0\bar{D}^0$ provide unique access to information about strong phase differences. Precision determination of the CKM phase $\gamma/\phi_3$ via $B \to D K$ decays depends upon constraints on charm mixing amplitudes, measurements of doubly-Cabibbo suppressed amplitudes and relative phases, and studies of correlated charmed meson decays tagged by flavor or CP eigenstates.  
CP-tagged $D^0 \to K^-\pi^+\pi^-\pi^+$ decays and CP-tagged $D^0 \to K^0_S\pi^+\pi^-$ Dalitz plots are only available at CLEO-c.
Using the 818~pb$^{-1}$ CLEO-c data sample produced by the Cornell Electron Storage Ring (CESR) at $\sqrt{s} = 3.77$ GeV, we perform analyses of these decays. We describe the techniques used to 
measure the $D$-decay parameters, and the CLEO-c impact on measurements of $\gamma/\phi_3$. 
\end{abstract}

\maketitle

\thispagestyle{fancy}


\section{INTRODUCTION}

\subsection{Measuring the CKM Phase $\gamma$}
Precision measurements of the weak phases that compose the unitarity triangle, $\alpha$, $\beta$ and $\gamma$, 
allow us to test the internal consistency of the Cabbibo-Kaboyashi-Maskawa (CKM) model and search for 
signatures of New Physics.  
The CKM phase $\gamma$ is only constrained by direct measurements to $({67 ^ { + 32} _ {- 25} })^{\circ}$ \cite{CKMFitter}. 
The most promising methods of determining the CKM phase
 $\gamma$ exploit the interference within $ B^{\pm}\!\to 
DK^{\pm}$ decays, where the neutral $D$ meson is a $ D^{0}$ or $ \bar{D}^{0}$. The most straightforward of these
strategies considers two-body final states of the $D$ meson, but additional information can be gained from %
strategies that consider multi-body final states.
The parameters 
associated with the specific 
final states needed for these analyses can be extracted from correlations within 
CLEO-c  \cite{CLEO} $\psi(3770)$ data. 

\subsection{Determination of the CKM phase $\gamma$ from $ B^{\pm}\!\to DK^{\pm}$}\label{sec:BDK}
The interference between decays of the type 
${B}^{\pm}\!\to {D}^{}{K}^{\pm}$ 
provide a theoretically clean method for extracting the CKM phase $\gamma$ when the ${D}^{0}$ and ${\bar{D}}^{0}$ mesons decay to a common final state, 
$f_{{D}}$. For example, we may write the ratio of the amplitudes between the suppressed amplitude and the dominant amplitude as:
\begin{equation}
\frac{
A({B}^{-}\!\rightarrow {\bar{D}}^{0} {K}^{-})}
{A({B}^{-}\!\rightarrow {D}^{0} {K}^{-})} = r_{B}e^{i(\delta_{B} - \gamma)},
\end{equation}
and we may write a similar ratio for ${B}^{+}\!\rightarrow {{D}}^{} {K}^{+}$.
The ratio of these amplitudes is a function of the ratio of the amplitudes' absolute magnitudes ($r_{B}$), a CP invariant strong phase 
difference ($\delta_{B}$), and the CKM weak phase $\gamma$. 
Due to color and CKM suppression, $r_{B} \sim 0.1$ \cite{CKMFitter}; therefore, the interference is generally small. A 
variety of strategies exist, however, that attempt to resolve this and maximize the achievable sensitivity to $\gamma$. 


\section{The ADS Formalism and $D \to K^-\pi^+$}
Atwood, Dunietz and Soni (ADS)\cite{ADS} have suggested considering $D$ decays to non-CP eigenstates as a way of maximizing sensitivity 
to $\gamma$. Final states such as $ K^{-}\pi^{+}$, which may arise from either a Cabibbo favored $D^{0}$ decay or a doubly 
Cabibbo suppressed $ \bar{D}^{0}$ decay, can lead to large interference effects and hence provide particular sensitivity to 
$\gamma$. 
This can be observed by considering the rates for the two possible $ B^{-}$ processes:
\begin{eqnarray}
\Gamma ({ B^{-}\!\to (K^{-}\pi^{+})_{D}K^{-}}) & \propto & 1  +  (r_B  r_D^{ K\pi})^2 
{}  {} +  2  r_B  r_D^{ K\pi}  \cos \left( \delta_B  -  \delta_D^{ K \pi}  -  \gamma \right),
\label{eq:fav1} \\
\Gamma ({ B^{-}\!\to (K^{+}\pi^{-})_{D}K^{-}}) & \propto & r_B^2  +  {(r_D^{ K\pi})}^2  
{}  {}  +  2  r_B  r_D^{ K\pi}  \cos \left( \delta_B  +  \delta_D^{ K \pi}  -  \gamma \right), 
\label{eq:dis1}
\end{eqnarray}
where $r_{D}^{ K\pi} = \sqrt{(0.3342 \pm 0.0084)\%}$ \cite{HFAG} parameterizes the relative suppression between 
$A_{{D}^{0}}$ and $A_{\bar{D}^{0}}$, and $\delta_D^{ K \pi}$, the relative strong phase 
difference.
By considering the other two rates associated with the $B^{+}$ decay, and combining this with information from 
decays to the CP-eigenstates $ K^{+}K^{-}$ and $\pi^{+}\pi^{-}$, an unambiguous determination of $\gamma$ can be made.
CLEO-c has recently measured $\delta_D^{ K \pi}$ to be $(22^{+14}_{-15})^{\circ}$ through a quantum correlated analysis of completely reconstructed $\psi(3770) \to D\bar{D}$ decays \cite{TQCA}. 

\section{$D \to K^-\pi^+\pi^+\pi^-$}

\subsection{Multi-body Extension to the ADS Method}
The ADS formalism can be extended by
considering multi-body decays of the $D$ meson. 
However, a multi-body $D$-decay amplitude is potentially 
different at any point within the decay phase space, because of the contribution of intermediate resonances. It is shown in 
Ref.~\cite{AS} how the rate equations for the two-body ADS method should be modified for use with multi-body final states. In the 
case of the $ B^{-}$ rates, for some inclusive final state $f$, Eq.~(\ref{eq:dis1}) becomes: 
\begin{equation}
\Gamma ({ B^{-}} \to (\bar{f})_{ D}{ K^{-}}) \propto \bar{A}_{f}^2 + r_B^{2}A_{f}^{2} + 2r_BR_{f}A_{f}\bar{A}_{f}\cos \left( 
\delta_B + \delta_D^{f} - \gamma \right), \label{eq:dis2}
\end{equation}
where $R_{f}$, the coherence factor, and $\delta_{D}^{f}$, the average strong phase difference, are defined as:
\begin{eqnarray}
A_{ f}^{2} & = & \int \vert A_{ D^{0}}(\mathbf x) \vert^{2}~d\mathbf x, 
~~~~\bar{A}_{ f}^{2} =  \int \vert A_{ \bar{D}^{0}}(\mathbf x) \vert^{2}~d\mathbf x, \\
R_{f}e^{i\delta_{ D}^{f}} & = & \frac{\int \vert A_{ D^{0}}(\mathbf x) \vert  \vert A_{ \bar{D}^{0}}(\mathbf x) \vert  
e^{i\zeta(\mathbf x)}~d\mathbf x}{A_{ f}  \bar{A}_{ f}} ~~~~\{ R_{f} \in {\mathbb R}~\vert~ 0 \leq R_{f} \leq 1 
\},~~\label{eq:Rf}
\end{eqnarray}
where $\mathbf x$ represents a point in multi-body phase space and $\zeta(\mathbf x)$ is the corresponding strong phase 
difference. 

\subsection{Determining $R_{f}$ and $\delta_{D}^{f}$ at CLEO-c}
\begin{figure*}[t]
\centering
\includegraphics[width=135mm]{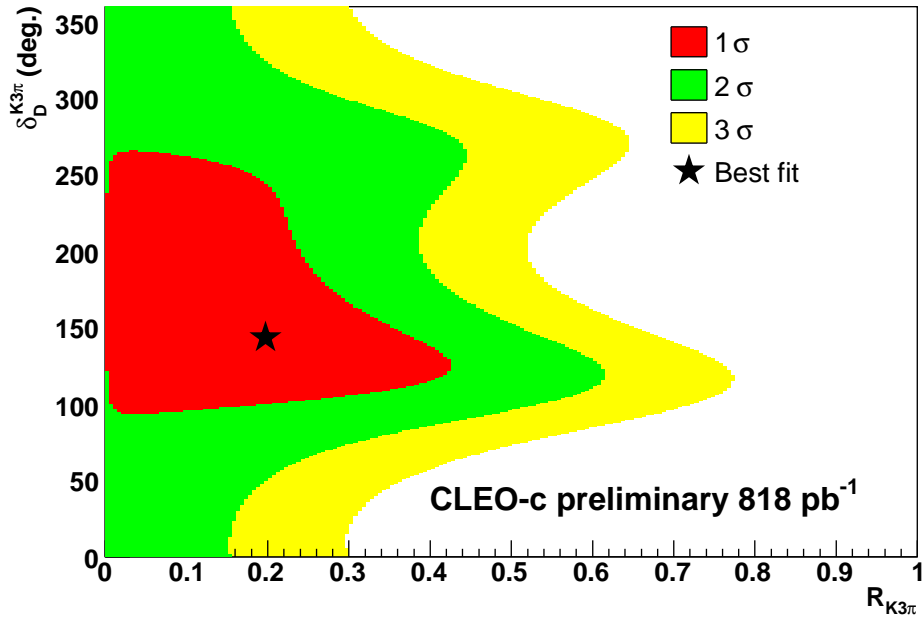}
\caption{(Preliminary) resulting limits on $R_{K3\pi}$ and $\delta^{K3\pi}_{D}$ at $1\sigma$, $2\sigma$ and $3\sigma$ levels. \label{fig:paramspace}}
\end{figure*}

It has 
been shown in Ref.~\cite{AS} that, double-tagged $ D^{0}\bar{D}^{0}$ rates measured at $\psi(3770)$ threshold provide sensitivity to 
both the coherence factor, $R_{f}$, and the average strong phase difference, $\delta_{D}^{f}$. Starting with the anti-symmetric 
wavefunction \cite{JR} of the $\psi(3770)$ and then calculating the matrix element for the general case of two inclusive final states, 
$F$ and $G$, the double-tagged rate is found to be proportional to:
\begin{equation}
\Gamma(F|G) \propto A_{F}^{2}  \bar{A}_{G}^{2}  +  \bar{A}_{F}^{2} A_{G}^{2} -  2  R_{F}  R_{G}  A_{F}  \bar{A}_{F} 
 A_{G}  \bar{A}_{G}  \cos(\delta_{D}^{F} - \delta_{D}^{G}).
\end{equation}
From this, one finds three separate cases of interest for accessing both the coherence factor and the average strong phase 
difference. These results are summarized in 
Ref. \cite{ANDREW}, where CLEO-c has provided a preliminary determination of   $R_{K3\pi}$ and $\delta^{K3\pi}_{D}$ for the instance of $F =  K\pi\pi\pi$ using  $818~{\rm pb}^{-1}$ of data taken at the $\psi(3770)$ resonance.
The resulting 
constraints on the parameters $R_{K3\pi}$ and $\delta^{K3\pi}_{D}$ from these preliminary measurements are shown in Fig.~\ref{fig:paramspace}.
It is apparent, from Fig.~\ref{fig:paramspace}, that the coherence across all phase space is low, reflecting 
the fact that many out of phase resonances contribute to the $K\pi\pi\pi$ final state. An inclusive analysis of this final state 
with the ADS analysis will therefore have low sensitivity to the phase $\gamma$, although the structure of Eq.~(\ref{eq:dis2}) makes 
it clear that such an analysis will allow for a determination of the amplitude ratio $r_{B}$, which is a very important auxiliary 
parameter in the $\gamma$ measurement.  

\begin{figure*}[t]
\centering
\includegraphics[width=135mm]{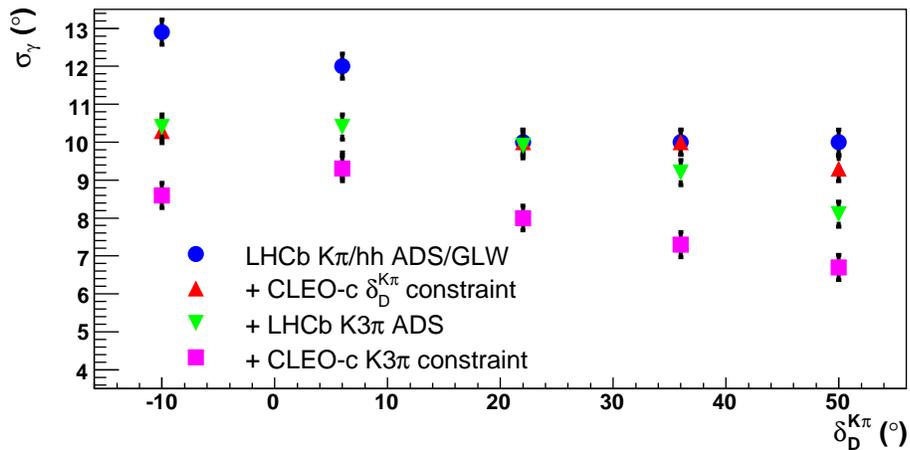}
\caption{Projections of the overall systematic uncertainty on $\gamma$ at LHC$b$, estimated for various values of of $\delta_D^{K\pi}$.} \label{DifferentCLEO-c}
\end{figure*}

Shown in Figure \ref{DifferentCLEO-c} are projections of the overall systematic uncertainty on $\gamma$ at LHC$b$ 
\cite{LHCbREF}. The figure demonstrates how the overall systematic uncertainty on $\gamma$ improves as additional information from CLEO-c is used in concert with expected LHC$b$ data samples documented in Reference \cite{lhcb2007004}.

\section{$D \to K_S\pi^+\pi^-$}

Dalitz plot analyses of the three-body decay ${D} \to K_S^0 \pi^+\pi^-$
together with studies of $B^{\pm}\to D K^{\pm}$ processes currently provide the best
measurements of the CKM weak phase $\gamma$ \cite{BabarNew, Belle}. 
However, ${D} \to K_S^0 \pi^+\pi^-$  Dalitz analyses are sensitive to the choice of the
model used to describe the three-body decay, which currently introduces a model systematic
uncertainty on the determination of  $\gamma$ which is greater than $5^{\circ}$ \cite{BabarNew}. 
For LHC$b$ and future Super-$B$ factories, this
uncertainty will become a major limitation.
A model independent approach to understanding the $D$ decay has been proposed by
Giri and further
investigated by Bondar \cite{GB}, which takes advantage of the quantum correlated $D^0/\bar D^0$
CLEO-c data produced at the $\psi(3770)$ resonance.

\begin{figure*}[t]
\centering
\includegraphics[width=135mm]{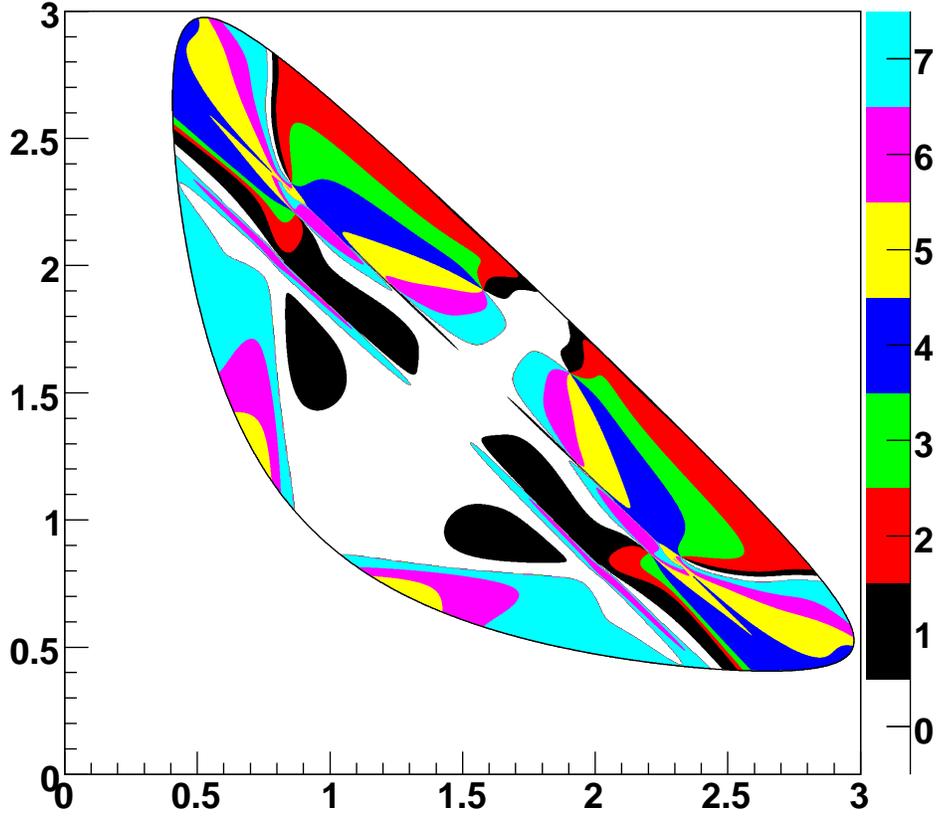}
\caption{Phase binning based on BaBar model. This plot is symmetric in $m_{K_S\pi^-}^2$ and $m_{K_S\pi^+}^2$} \label{Babar}
\end{figure*}

Consider a Dalitz plot in which we define $x = m_{K_S\pi^-}^2$ and $y = m_{K_S\pi^+}^2$. Both $D^0 \to K_S^0 \pi^+\pi^-$ and $\bar{D}^0 \to K_S^0 \pi^+\pi^-$ decays appear on this plot. We then divide the Dalitz plot into regions which are expected to have about the same relative strong phase difference between the $D^0$ and $\bar{D}^0$ decays, based on the $D^0 \to K_S^0 \pi^+\pi^-$ decay model from BaBar \cite{BabarModel}, as shown in Figure \ref{Babar}. 
Assuming the amplitude for the $D^0 \to K_S^0 \pi^+\pi^-$  process is $f_D(x,y)$, we can define the bin-averaged cosine, $c_i$, and bin-averaged sine, $s_i$, for each bin $i$ as follows:
\begin{eqnarray}
f_D(x,y) &=& |f_D(x,y)|e^{i\delta_D(x,y)} \\
c_i &=& \frac{1}{\sqrt{F_iF_{\bar\imath}}}\int_{D_i}|f_D(x,y)||f_D(y,x)| 
{\cos(\delta_{x,y}-\delta_{y,x})}dxdy\\
s_i &=& \frac{1}{\sqrt{F_iF_{\bar\imath}}}\int_{D_i}|f_D(x,y)||f_D(y,x)| 
{\sin(\delta_{x,y}-\delta_{y,x})}dxdy
\end{eqnarray}

Using the 818 pb$^{-1}$ $\psi(3770) \to D^0 \bar{D}^0$ data sample collected by CLEO-c, we can measure the strong phase parameters, $c_i$ and $s_i$, using fully reconstructed
$D^0 \bar{D}^0$ pairs with $K_S^0 \pi^+\pi^-$ vs. flavor states, CP eigenstates, and
double $K_S^0 \pi^+\pi^-$ samples. 

\begin{figure*}[t]
\centering
(a)\includegraphics[width= 135mm]{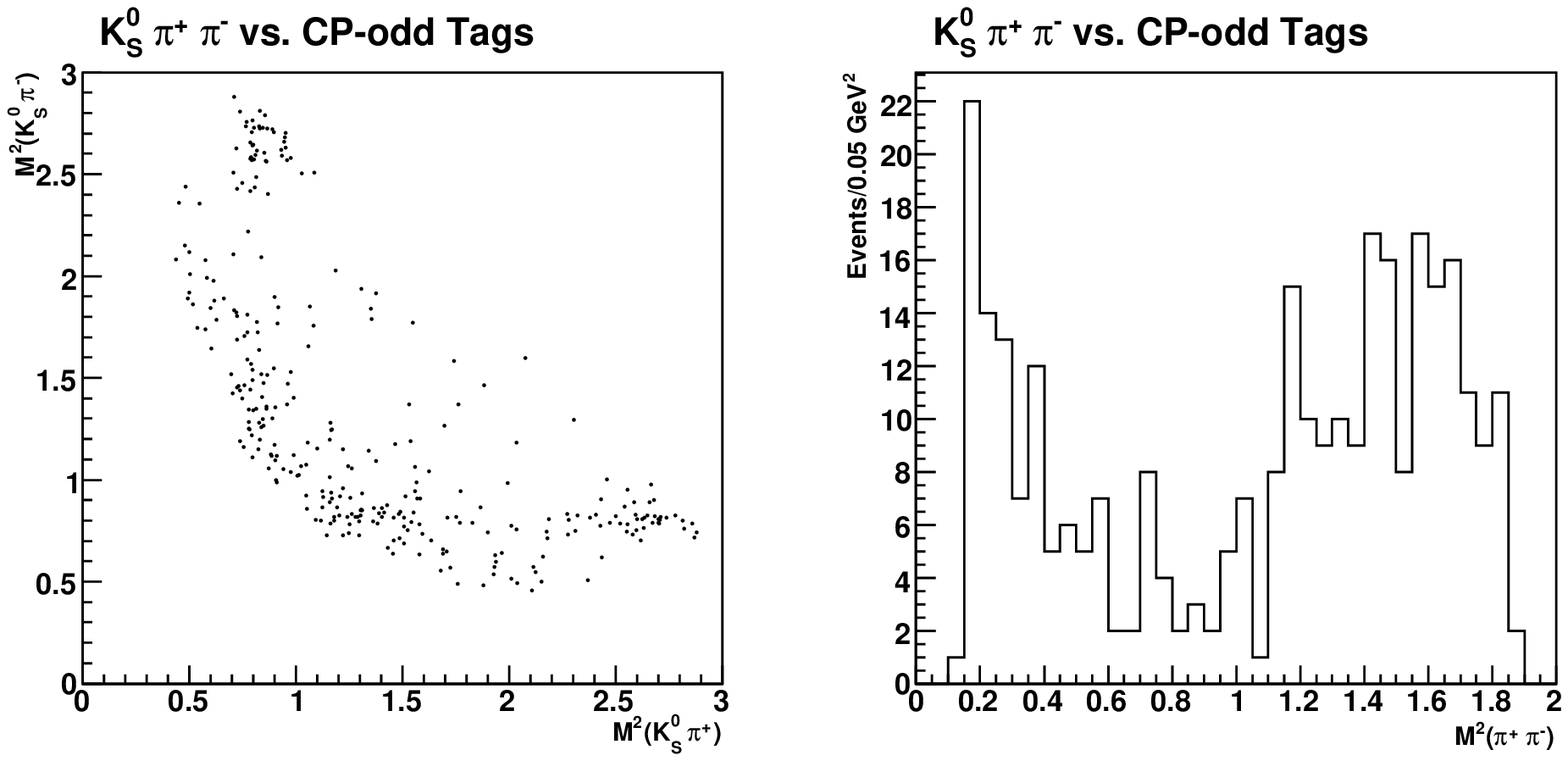}\\
(b)\includegraphics[width= 135mm]{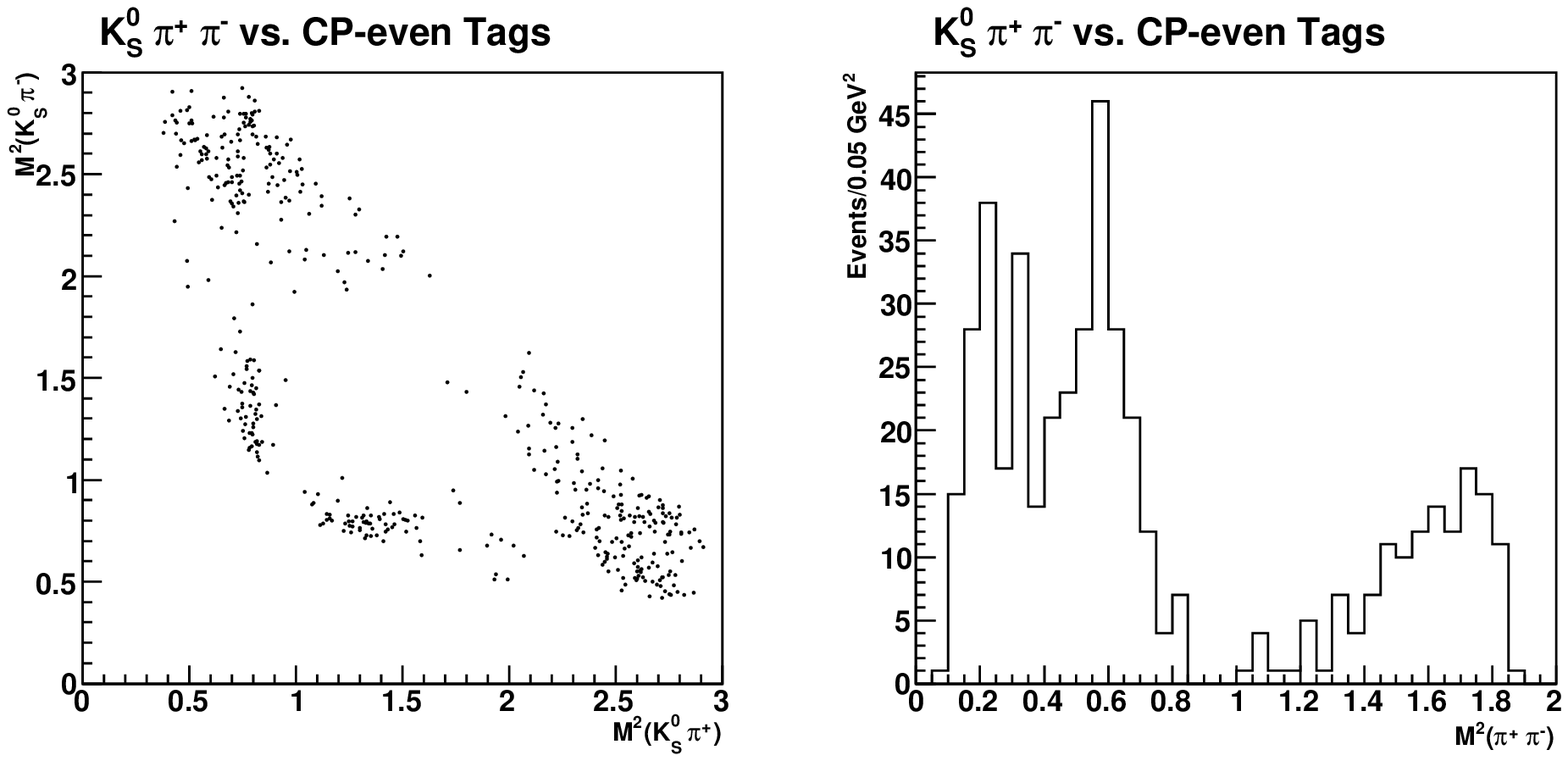}
\caption{(a) CP-odd Tags. (b) CP-even Tags.} \label{CPTags}
\end{figure*}

We may create a CP-tagged sample $K_S^0 \pi^+\pi^-$ events by requiring the neutral $D$ which does not decay to $K_S^0 \pi^+\pi^-$ to decay to states of definite CP ($\pi^+\pi^-, K^+K^-, K^0_S\pi^0\pi^0, K^0_L\pi^0, K^0_S \pi^0, K^0_S \eta,~\rm{and}~K^0_S \omega$). The CP-tagged $K_S^0 \pi^+\pi^-$ sample only allows us to measure $c_i$, and not $s_i$, in each bin. 
It can be shown that the bin averaged cosine in each of these bins is:
\begin{eqnarray}
c_i = \frac{(M_i^+/S_+-M_i^-/S_-)}{(M_i^+/S_++M_i^-/S_-)}
\frac{(K_i+K_{\bar\imath})}{2\sqrt{K_iK_{\bar\imath}}},
\end{eqnarray}
where  $M_i^+(M_i^-)$ is the number of CP even(odd)-tagged $K_S^0 \pi^+ \pi^-$ events in each bin
 and  $K_i(K_{\bar\imath})$ is the number of flavor tagged $K_S^0 \pi^+ \pi^-$ events in each bin. In our analysis, we use hadronic flavor-tags ($K^- \pi^+, K^- \pi^+ \pi^0, ~\rm{and}~K^- \pi^+ \pi^+ \pi^-$), for which doubly-Cabbibo suppressed decays are considered in evaluation of the systematic error.
There are $\sim$ 800 CP-tagged events in the sample we use to determine $c_i$ in each bin, which are shown in Figure \ref{CPTags}.

Using the $K_S^0 \pi^+ \pi^-$ vs. $K_S^0 \pi^+ \pi^-$ sample, one can extract 
$c_i$ and $s_i$ simultaneouly. The number of double-tagged events $M_{i,j}$ can be related to the number of flavor-tags for each $D$ decay: 
\begin{eqnarray}
M_{i,j}=\frac{1}{2N_{D,\bar D}\mathcal B_f^2}(K_iK_{\bar\jmath}+K_{\bar\imath}K_j-2\sqrt{K_iK_{\bar\jmath}K_{\bar\imath}K_j}(c_ic_j+s_is_j)),
\end{eqnarray}
where $\mathcal{B}_f$ is the branching ratio of $K_S^0 \pi^+ \pi^-$, and $N_{D,\bar D}$ is the number of $\psi(3770)$ decays, assuming 100\% efficiency.
There are $\sim$ 450  $K_S^0 \pi^+ \pi^-$ vs. $K_S^0 \pi^+ \pi^-$  events in the sample of  $K_S^0 \pi^+ \pi^-$ vs. $K_S^0 \pi^+ \pi^-$ events.

The latest preliminary CLEO results for $c_i$ and $s_i$ from both $K_S^0 \pi^+ \pi^-$ CP-Tags and $K_S^0 \pi^+ \pi^-$ vs. $K_S^0 \pi^+ \pi^-$ events are shown in Table~\ref{tab:cisi}.
\begin{table}[t]
\centering
\label{tab:cisi}
\begin{tabular}{|c|r|r|r|}        
\hline
i     & $c_i~(K_S^0 \pi^+\pi^-$~{vs.~CP-Tags}$)$& $c_i~(K_S^0 \pi^+\pi^-~\rm{vs.}~K_S^0 \pi^+\pi^-$) & $s_i~(K_S^0 \pi^+\pi^-~\rm{vs.}~K_S^0 \pi^+\pi^-)$ \\
\hline
1&0.706 $\pm$ 0.069 $\pm$ 0.028&  0.779 $\pm$ 0.087 $\pm$ 0.062 & 0.380 $\pm$ 0.179 $\pm$ 0.085 \\
2 &0.586 $\pm$ 0.126 $\pm$ 0.037&  0.874 $\pm$ 0.120 $\pm$ 0.113 &  0.137 $\pm$ 0.260 $\pm$ 0.084 \\
3 &0.041 $\pm$ 0.120 $\pm$ 0.043 &0.003 $\pm$ 0.166 $\pm$ 0.152 & 0.749 $\pm$ 0.145 $\pm$ 0.053 \\
4 &-0.510 $\pm$ 0.178 $\pm$ 0.074&-0.165 $\pm$ 0.323 $\pm$ 0.152 & 0.490 $\pm$ 0.400 $\pm$ 0.093 \\
5 &-0.949 $\pm$ 0.063 $\pm$ 0.029 &-0.929 $\pm$ 0.058 $\pm$ 0.044 & 0.141 $\pm$ 0.268 $\pm$ 0.085 \\
6 &-0.807 $\pm$ 0.108 $\pm$ 0.039 &-0.472 $\pm$ 0.196 $\pm$ 0.099 & -0.679 $\pm$ 0.203 $\pm$ 0.059 \\
7 &0.085 $\pm$ 0.154 $\pm$ 0.046&0.459 $\pm$ 0.204 $\pm$ 0.170 & -0.558 $\pm$ 0.367 $\pm$ 0.106 \\
8 &0.339 $\pm$ 0.082 $\pm$ 0.024&0.526 $\pm$ 0.109 $\pm$ 0.114 & -0.376 $\pm$ 0.169 $\pm$ 0.060 \\
\hline
\end{tabular}
\caption{Preliminary CLEO results for $c_i$ and $s_i$ with respect to a particular type of tag.}
\end{table}

With the measurements presented here, the systematic uncertainty resulting from our understanding of the $D$ decays is lowered 
to $\sim2^{\circ}$, which is calculated using the methods reported in Reference \cite{lhcb2007141}.

\begin{acknowledgments}
The author wishes to thank David Asner, Neville Harnew, Qing He, Jim Libby, Andrew Powell, Jonas Rademacker, Ed Thorndike, and Guy Wilkinson for their contributions and guidance in preparing these results.
\end{acknowledgments}


\end{document}